\documentclass[11pt]{article}
\usepackage{graphicx,float,pict2e}
\usepackage[table]{xcolor}
\usepackage[normalem]{ulem}
\usepackage{graphicx}
\usepackage{subfig}
\usepackage{booktabs}
\usepackage{epsfig}
\usepackage{color}
\usepackage{rotating}
\usepackage{verbatim}
\usepackage{hyperref}
\usepackage{amsmath}
\usepackage{amssymb}
\usepackage{amsfonts}
\usepackage{authblk}
\title{Extreme-$K$ categorical samples problem}

\author[1, 3]{\small Chou, Elizabeth}
\author[2, 3]{\small McVey, Catie}
\author[4]{\small Hsieh, Yin-Chen}
\author[5]{\small  Enriquez, Sabrina}
\author[6 *]{\small Hsieh, Fushing}

\affil[1]{\footnotesize Department of Statistics, National Chengchi University, Taiwan.}
\affil[2]{\footnotesize Department of Animal Science, University of California at Davis, CA, 95616.}
\affil[3]{\footnotesize First two authors contribute equally.}
\affil[4]{\footnotesize Dept of BioSciences, Fisheries, \& Economics UiT Arctic University of Norway, Troms${\o}$, Norway.}
\affil[5]{\footnotesize Department of Mathematics, University of California at Davis, CA, 95616.}
\affil[6]{\footnotesize Department of Statistics, University of California at Davis, CA, 95616.}
\affil[*]{Corresponding author: Hsieh, Fushing. E-mail: fhsieh@ucdavis.edu}

\date{}
\begin{document}
\maketitle

\begin{abstract}
With histograms as its foundation, we develop Categorical Exploratory Data Analysis (CEDA) under the extreme-$K$ sample problem, and illustrate its universal applicability through four 1D categorical datasets. Given a sizable $K$, CEDA's ultimate goal amounts to discover by data's information content via carrying out two data-driven computational tasks: 1) establish a tree geometry upon $K$ populations as a platform for discovering a wide spectrum of patterns among populations; 2) evaluate each geometric pattern's reliability. In CEDA developments, each population gives rise to a row vector of categories proportions. Upon the data matrix's row-axis, we discuss the pros and cons of Euclidean distance against its weighted version for building a binary clustering tree geometry. The criterion of choice rests on degrees of uniformness in column-blocks framed by this binary clustering tree.  Each tree-leaf (population) is then encoded with a binary code sequence, so is tree-based pattern. For evaluating reliability, we adopt row-wise multinomial randomness to generate an ensemble of matrix mimicries, so an ensemble of mimicked binary trees.  Reliability of any observed pattern is its recurrence rate within the tree ensemble. A high reliability value means a deterministic pattern. Our four applications of CEDA illuminate four significant aspects of extreme-$K$ sample problems.
\end{abstract}


\section{Introduction}
Categorical data is ubiquitous in every corner business and industry, the sciences, and the humanities. It is fair to say that categorical data is no less prevalent than the other data types: continuous or discrete. Indeed, in some fields, categorical data can be considered the predominant data type. Nonetheless, the popularity of Categorical Data Analysis, as a topic of statistics, is disproportional to the extreme. In fact, this topic is indeed disappearing in recent decades under the shadows of Machine Learning and Data Science. The following two facts offer a glimpse into its current state of affairs. 1) It is no longer a required course for M.A or PhD students in Statistics. 2) It hardly appears in titles in our Statistics department seminars. If you ask students how would they deal with a categorical variable in data analysis, then, with a high probability, their answers are exactly the same: making use of dummy binary variables. But is this surrogate representation of a categorical variable legitimate in data analysis?

Since a categorical variable does not naturally lend itself to the fundamental statistical concepts of mean and variance, neither Regression analysis nor Analysis of Variance (ANOVA) are directly or intuitively applied. Traditional techniques for categorical data analysis have mainly focused on associations between two or three categorical variables within a population via logit model and the like \cite{goodman, fienberg}. Such methodologies are limited exclusively to the domain of Categorical Data analysis. It may seem striking to argue that categorical variable-based analysis is a new fundamental frontier of data analysis in Data Science, but we endeavor to offer glimpse of this perspective in this paper.

Consider one categorical variable with data sampled from K populations with K being large. This data setting is termed "Extreme-K categorical sample problem". Our proposed computational paradigm, called categorical exploratory data analysis (CEDA), is nonparametric and exploratory in nature. Since our methodology is based on K histograms, it may accommodate all data-types. That is, each continuous variable can be effectively converted into a categorical one because its distribution function can be approximated very well by a piecewise-linear version, which is a distribution of a possibly-gapped histogram \cite{hsiehroy}. As a histogram is universal for all variables, our CEDA is also universally applicable.

Further, our categorical exploratory data analysis (CEDA) is developed to achieve the universal goal of data analysis: extracting the data's full information content. This goal is particularly relevant when the number (K) of populations involved is large. On one hand, the system containing such a collection of populations is likely vast and complex. On the other hand, researchers interested in such a system are unlikely to have complete a priori knowledge what hypotheses may be of interest and what hypotheses need not apply to the system at hand. Hence, it becomes essential that exploratory data analysis is conducted in hopes revealing all potential patterns that might be truly sustained by data. Ideally our CEDA would render these patterns and arrange them on a geometry that reveals both the global and local structures on the space of populations. Upon such a geometry, researchers can more efficiently invest further analytical efforts. Thus, it is worth keeping in mind that when patterns are to be discovered in a complex system, hypotheses determined in a priori fashion are neither realistic nor practical.

A tree geometry on a space of populations, in general, is an informative construct that accommodates a fairly wide scope of patterns that are commonly found to be sustained by data. This capacity is afforded by the tree's sequential bifurcating structures, which yield a well-guided map for making pairwise comparisons. More importantly, a tree gives rise to multiscale maps of clustering compositions that will allow researchers to discover diverse patterns of groupings for populations of various sizes, as well as patterns of separations among groups. Such patterns must ultimately be confirmed or disputed via evaluations of their uncertainty or reliability, but their discoveries upon a tree's branching geometry are the first step in the critical process of revealing hidden knowledge and intelligence via data analysis.

Our focus on a tree's geometry doesn't imply that we rule out the possibility of vital patterns existing beyond a tree geometry. Our focal clustering tree in fact is the natural marginal statistics of the observed bipartite network represented by the observed data matrix. A tree of this nature is specifically called a label-embedding tree in Multiclass Classification (MCC), which is a major machine learning topic \cite{hsiehwang,hsiehchou}. Although the extreme-K sample problem is a special case of MCC involving only one feature, the classification task is hardly its primary focus. This drastic shift on presumed primary tasks within MCC, due simply to differences in number of involved features, strongly indicates that the classification task by-and-large is neither universal nor fundamental in MCC. In contrast, the task of extracting full information content contained in data is universal and fundamental. Here a population embedding tree is only one component of information content.

The data sampled from K populations indeed contains other components of information content that are key to our understanding of the system of interest. Our CEDA also develops a mimicking algorithm to evaluate the uncertainty or reliability of each structural pattern of interest observed on a population embedding tree. This mimicking algorithm generates a large ensemble of mimicries of the observed data matrix, with which we generate a large ensemble of mimicries of a population embedding tree. Given that each mimicry replicates as nearly as possible the full stochasticity embraced by the original data matrix, this ensemble of mimicked trees is designed to supposedly disentangle deterministic structures from the stochastic structures of the original tree. This separation of deterministic and stochastic features of the observed tree is an essential and necessary component of extracting the data's full information content.

Since this ensemble of mimicries can have an arbitrarily large size, the heterogeneous uncertainties or reliability for all possible perceivable patterns are probabilities in reality. Further, they become visible and explainable when we superimpose the original tree onto the data matrix's row-axis. This rearranged data matrix is termed a heatmap. Upon such a heatmap, all large and small tree branches are seen being attached with a series of column-blocks with various degrees of uniformness. In other words, we are able to elucidate why some observed tree patterns are confirmed as deterministic, because they can't be swayed by the data's stochasticity, whereas some tree patterns are declared as stochastic, because they can be swayed by data's stochasticity. These collections of visible and readable pattern information are the data's chief information content.

When K is large, each clustering tree based on the original or a mimicked data matrix is likely geometrically complex. However, because of their bifurcating nature, each population-leaf can be represented (encoded) with a binary code sequence for its location on a tree. Hence, each observable pattern upon a tree geometry can be located precisely via a binary code segment. That is, each tree geometry based on the observed or a mimicked data matrix will give rise to a table of population-leaf-specific binary code sequences. Consequently, data's information content can be efficiently extracted by algorithmically working upon the collection of tables of binary code sequences. Such information content is only possible and feasible by virtue of modern computational resources. Based on such information content, an Extreme-K sample problem's data matrix can be made into an A.I.

In this paper, we analyze four data sets: 1) Country of origin of foreign students attending national universities in Taiwan; 2) Amino acid content of proteins; 3) Behavioral time budgets of organic dairy cows; and 4) Major League Baseball (MLB) pitchers' pitch-types.

\section{Structural details in an extreme-$K$ categorical sample problem.}
Let the focal categorical variable commonly realize its qualitative categories $\{J_1, J_2, .....J_M\}$ across all $K$ populations with varying total counts $\{N_k\}^K_{k=1}$. We turn the observed category-counts into proportions denoted as $p^o_{km}$ for all $k=1,...,K$ and $m=1,...,M$. That is, we observe a $M$-vector $P^o_k=(p^o_{k1},...,p^o_{km},...p^o_{kM})$, which is also a histogram, from the $k$-th population. What then is information content contained in $\{P^o_k\}^K_{k=1}$ How should we build a tree geometry upon the $K$ populations based on the $\{P^o_k\}^K_{k=1}$? These are fundamental questions that need to be answered.  In Statistics, due to the largeness of K and categorical nature of the data, such questions do not fall within the domain of classic K-sample problem.

For simplicity, we assume independency among all $K$ observed histograms, and each histogram is resulted from the identically independently sampling within each population. These assumptions are operational ones. They might be violated even for all examples considered here. Then, given independency, the $k$-th population's stochastic and deterministic structures are coherently captured by Multinomial random variables $MN(N_k,P^o_k)$ with  $k=1,...,K$. Therefore, we can simulate $N_k$ categorical samples from $MN(N_k,P^o_k)$ to make up a mimicry of $P^o_k$, denoted as $P^b_k=(p^b_{k1},....., p^b_{kM})$. Let's create a large mimicry ensemble $\{\{ P^b_k \}^K_{k=1}\}^B_{b=1}$ in accordance with the independency assumptions imposed on the original data $\{P^o_k\}^K_{k=1}$. We compile all population-vectors into a $K \times M$ matrix format:
\[
{\cal P}^o=[P^o_{km}].
\]
This matrix ${\cal P}^o$ is exactly the compositional view of the ``histogram'' of pooled data. On the other hand, it is a bipartite network. Likewise, we denote an ensemble of matrix-mimicries as \[
\{{\cal P}^b\}^B_{b=1}=\{\{(p^b_{k1},....., p^b_{kM})\}^K_{k=1}\}^B_{b=1}.
\]

Next, if we take the $K$ populations to constitute a system under study that gives rise to the data matrix ${\cal P}^o$, then the $M$-categories being arranged along the column axis are legitimate $M$-features. In spite of the constraint of $\sum^M_{m=1} p^o_{km}=1$, these features' associative patterns ideally would reflect the system's hidden nature that we seek to computationally recover.

A glimpse of such associative patterns can be visualized in the following manner. We can build a possibly gapped histogram based on each column of ${\cal P}^o$ \cite{hsiehroy}.  Then we accordingly build a contingency table for each pair of columns and calculate its mutual conditional entropy to assess its degree of association \cite{hsiehshanyu}. The lower mutual conditional entropy value is, the higher the association between the pair of columns (or categories). We can encode all bins of the two marginal histograms with distinct colors. In this fashion, a population is represented by a bivariate color-codes. Then associative patterns from this highly associative pair of columns can then be visualized via a properly permuted $K \times 2$ color-coded matrix, which is carried out by simply grouping all of the same bivariate color-codes together. This permuted matrix reveals associative patterns because there will be only a small number of distinct bivariate codes present due to their high association. Such associative patterns under a slightly different color-coding scheme can be seen in Fig.~\ref{fig:cow1} and Fig.~\ref{fig:cow4} and many others figures reported afterward.

Likewise we can put together three highly associated columns to reveal their association patterns, and so on. Within this protocol of demonstrating associative patterns among category-based columns or features, we need to choose the number of bins in individual histograms and then carry out grouping on multivariate color-code. Here we appeal to a tree geometry to resolve all aforementioned tasks in an automatic and systematic fashion, since its bifurcating structures are to be dictated by multiscale associative patterns underlying the $M$ features. This fact can be explicitly seen. When we superimpose a computed tree geometry upon the row-axis of data matrix ${\cal P}^o$, we expect to see that various levels of tree branches indeed frame out multiple column-blocks with various degrees of uniformness within each of the $M$ columns, as would be seen all of the real extreme-$K$sample problems considered here.

The capability of a tree geometry in revealing multiscale associative patterns for extreme-$K$ sample problem is in fact blurred by the stochasticity of all orders of dependence. For example, pairwise correlations and higher order correlations, existing among the $M$ components of vector $P^o_k=(p^o_{k1},...,p^o_{km},...p^o_{kM})$ within each population can blur the patterns we seek to discover. Under the Multinomial random variable $MN(n_k,P^o_k)$, each mimicry $P^b_k=(p^b_{k1},....., p^b_{kM})$ indeed replicates such pairwise correlations, as listed below:
\begin{eqnarray*}
Cov[p^b_{km},p^b_{km'}]&=&E[p^b_{km}p^b_{km'}]-p^o_{km}p^o_{km'}\\
&=&\frac{-1}{N_k}p^o_{km}p^o_{km'}, {\textrm{if}}\; m\neq m';\\
Var[p^b_{km}]&=&\frac{1}{N_k}p^o_{km}(1-p^o_{km}), {\textrm{if}} \; m = m'.
\end{eqnarray*}

Thus, in each of the real examples we consider, our patterns also get blurred in the tree geom. But the unknown and potentially versatile higher order dependence among the $M$ components of vector $P^o_k=(p^o_{k1},...,p^o_{km},...p^o_{kM})$ are surely lost due to employing identically independent sampling when carrying out mimicry $P^b_k=(p^b_{k1},....., p^b_{kM})$. We explicitly demonstrate such losses in the cow's time budget example. How such losses could impact our discovering associative patterns among the $M$-features is still unknown at this stage.

The above formulas of within-population correlations and variances also reveal that the heterogeneous sample sizes $\{N_k\}^K_{k=1}$ lead to the heterogeneous precisions of $\{p^o_{km}\}^K_{k=1}$ within a bin-locality. The small $N_k$s likely further enhance the blurring effect in uncovering the associative patterns of $M$-features via the concept of uncertainty or reliability.

\section{How should a tree geometry be computed?}
Our focal tree geometry is a clustering tree based on the Hierarchical Clustering (HC) algorithm. It groups similar $M$-vectors together, and at the same time bifurcates dissimilar groups into nearby or faraway branches according to the degrees of their mutual dissimilarity. The HC algorithm operates upon a chosen distance measure calculated between two $M$ vectors, and the module of distance computed between two sets of vectors. We decide to use the module called Ward.D2 for its stability. The resultant tree geometry is proven to be more sensitive to the choice of a distance measure than the choice of module.

Two distance measures are considered here. One is the Euclidean distance $d_0(.,.)$ in $R^M$. This distance measure employs equal weights across all its $M$ components when evaluating dissimilarity among the $K$ row-vectors of data matrix ${\cal P}^o$. This constant weighting scheme is robust. But it is insensitive to the heterogeneous precisions due to the heterogeneity of $\{N_k\}^K_{k=1}$. Such insensitivity could cause biases in measuring distances.

In contrast, a version of fine-tuned Euclidean distance is proposed as follows. By knowing the component-wise variations for all localities at $m$-th category and $k$-th population across $m=1,... M$ and $k=1,..,K$, it seems reasonable to modify the popularly employed Euclidean distance in $R^M$ among $\{P^o_k\}^K_{k=1}$ and $\{P^b_k\}^K_{k=1}\}$ for all $b=1,..., B$ according to the locality variations evaluated and denoted as: given the $B$ can be arbitrary large,

\[
Var[p^b_{km}]=\frac{1}{N_k}p^o_{km}(1-p^o_{km})\approx \frac{1}{B}\sum^B_{b=1}(p^o_{km}-p^b_{km})^2.
\]
Here $ Var[p^b_{km}]$ is the mimicking variation of $p^b_{km}$ for all $k=1,.., K$ and $m=1,.... M$. Then we have
\[
Var[p^b_{km}-p^b_{k'm}]=Var[p^b_{km}]+Var[p^b_{k'm}].
\]
Then one version of fine tuning on the Euclidean distance for observed and mimicked vectors should be:
\begin{eqnarray*}
d^*(P^o_k, P^o_{k'})&=&\sum^M_{m=1}\frac{[p^o_{km}- p^o_{k'm}]^2}{Var[p^b_{km}]+Var[p^b_{k'm}]}\\
d^*(P^b_k, P^b_{k'})&=&\sum^M_{m=1}\frac{[p^b_{km}- p^b_{k'm}]^2}{Var[p^b_{km}]+Var[p^b_{k'm}]}
\end{eqnarray*}

The key rationale underlying this re-scaled fashion is given follows. Each component-wise discrepancy is distributed as: if $N_k$ is large enough, then
\[
\frac{[p^b_{km}- p^b_{k'm}]^2}{Var[p^b_{km}]+Var[p^b_{k'm}]}\sim \frac{[p^o_{km}- p^o_{k'm}]^2}{Var[p^b_{km}]+Var[p^b_{k'm}]}+ \chi^2_1.
\]
where $\sim$ means ``distributed as'', $\chi^2_1$ denotes Chi-square distribution with degree freedom 1 and $\frac{[p^o_{km}- p^o_{k'm}]^2}{Var[p^b_{km}]+Var[p^b_{k'm}]}$ is a non-centrality .

Therefore, for all $k$ and $k'$ across all $b$, we have:
\[
d^*(P^b_k, P^b_{k'})\sim d^*(P^o_k, P^o_{k'})+ \chi^2_M.
\]

When all sample sizes $\{N_k\}^K_{k=1}$ are uniformly large enough, then this expression of $ d^*(P^b_k, P^b_{k'})$ ensures that all involved distances are comparable across all $K$ row-vectors. But it is not the case for Euclidean distances $ d_0(P^b_k, P^b_{k'})$ due to heterogeneity of $\{N_k\}^K_{k=1}$.

The above equality implies that the HC tree geometry, denoted as  $\tilde{T}^o$,  computed by applying the HC algorithm based the $K\times K$ distance matrix $[d^*(P^o_k, P^o_{k'})]$ ($1\leq k, k' \leq K$) should be ``stochastically homomorphic'' with tree geometries $\{\tilde{T}^b\}^B_{b=1}$ associating with $K\times K$ distance matrix $[d^*(P^b_k, P^b_{k'})]$ for all $1 \leq b \leq B$. In this fashion, we accommodate locality variations across all categories, on one hand. And we make the component-wise discrepancies comparable across all $M$ categories as well as across all $K$ populations. This is the true spirit of doing re-scaling. This is the essential concept of Analysis of Variance (ANOVA).

However, when some of $\{N_k\}^K_{k=1}$ are small, then the re-scaling could be destabilized. The resultant tree geometry ${\tilde T}^o$ would end up contain many unreliable structures. In such a case, the tree geometry based on Euclidean distance $d_0(.,.)$ might turn out more sensible.

\section{A real data illustration}
We illustrate our CEDA and reliability computations using behavioral time budget data collected from a herd of organic dairy cows \cite{manriquez}. Via this data set, we demonstrate the benefits of adopting $d^*(.,.)$ over $d_o(.,.)$ in our associative pattern explorations. We also use this data to demonstrate the benefits of knowing how exactly the identically independently sampling assumption is violated.

Data for these analyses were generated as part of a comprehensive feed trial conducted by the Dairy Systems Group at Colorado State University. After a herd of 200 cows had been fully established, a commercial ear tag accelerometer device was used to automatically record the behaviors of each animal. producing each hour a breakdown of total minutes a cow was engaged in five behaviors: active, highly active, nonactive, ruminating, or eating. After scrubbing data for cows with sensor-receiver failures and health complications, complete records were available from 124 cows over a period of 42 days. For this analysis, records were aggregated into an overall time budget spanning all 42 days; however, time budgets could have also been generated at the hourly or daily level. Due to daily schedule changes for managemental reasons, the identically independently sampling doesn't hold across the entire temporal span, which for each cow is about 1008 hours (60060 min). With such uniformly large data sizes for all involved cows, we expect that the distance $d^*(.,.)$ would lead us to more evident patterns than $d_o(.,.)$ could.

\begin{figure*}[t]
		\centering
		\includegraphics[width=4.7in]{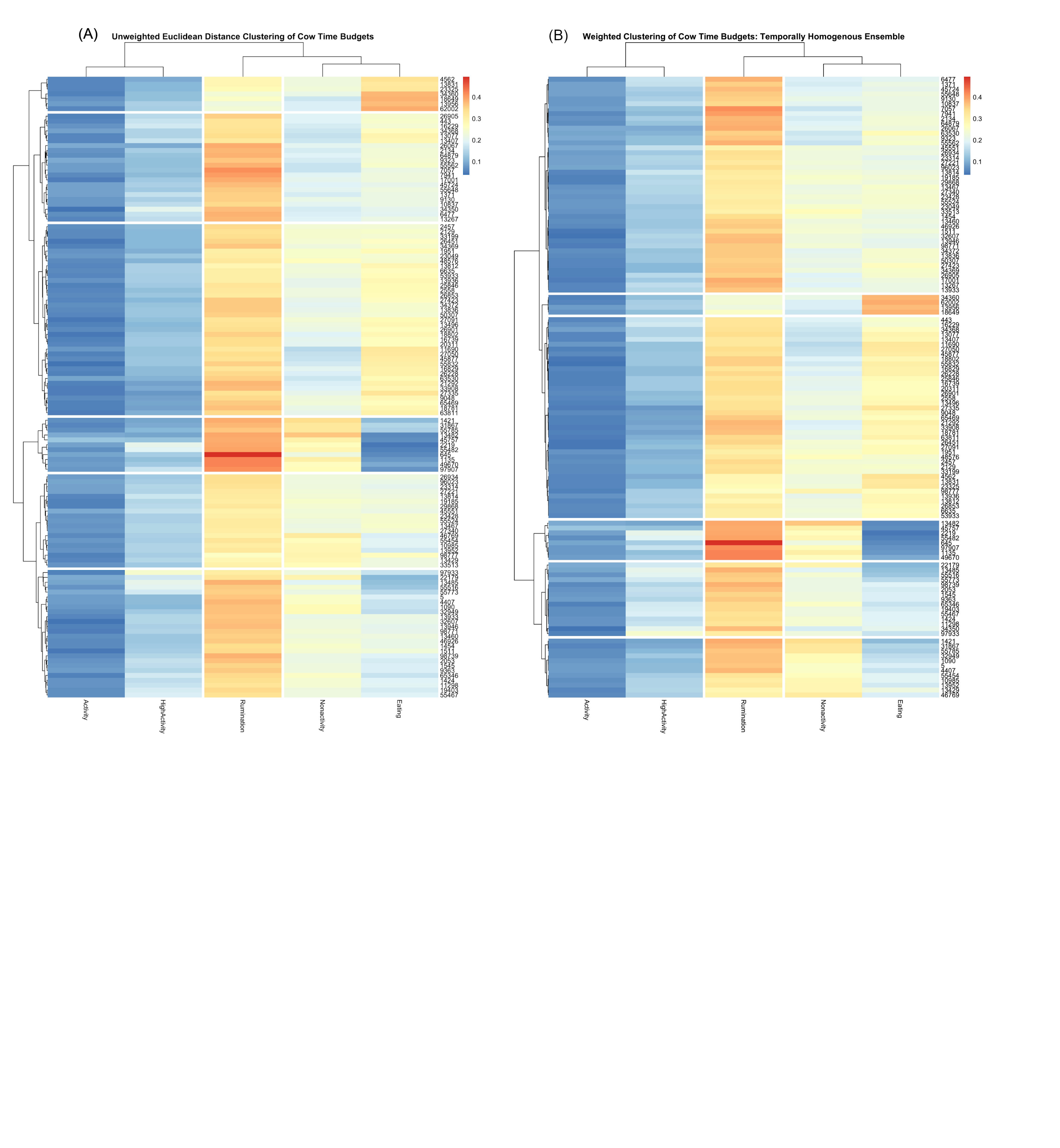}
		\caption{Two heatmaps of cows' time budgets superimposed with a HC-tree based on distance measure (A) $d_o(.,.)$; (B) $d^*(.,.)$.}
		\label{fig:cow1}
	\end{figure*}

For illustrating purpose, we demonstrate this evident fact by comparing the two heatmaps based on the data matrix of 124 cows, as shown in Fig.~\ref{fig:cow1}. We immediately observe that panel (B) reveals clearer uniformness upon the 6 cluster-based column-blocks across all 5 columns of behaviors, particularly on the columns of nonactive and eating. (If it is necessary, then we could also use within-cluster total-variations to quantify such a comparison.) Since similarity is dictated by the employed distance measure, such clearer associative patterns with uniformness imply that cows sharing the same cluster are indeed truly similar to each other. Therefore, we conclude that $d^*(.,.)$ indeed brings out more solid patterns than $d_o(.,.)$ could in this example.

Next, we turn to the the benefits realized in relaxing the identically independently sampling assumption. Under the temporally homogeneous assumption, the stochasticity captured by aggregating 42-days time budget via a Multinomial random variable $MN(N_k,P_k)$ doesn't reflect the reality of temporal heterogeneity in the observed data matrix ${\cal P}^o$ due to within- and between- day changes in cow management schedules. It is curious as well as interesting to see what differences could possibly be made if we build the temporal heterogeneity into the mimicking protocol used for generating the ensemble of matrix-mimicries $\{{\cal P}^b\}^B_{b=1}$. That is, we generate hourly time-budgets across 42-days for each cow by sampling from the multinomial at each observation hour, and then pool the data together to form an aggregate 42-days time budget for each cow.

\begin{figure*}[t]
		\centering
		\includegraphics[width=6in]{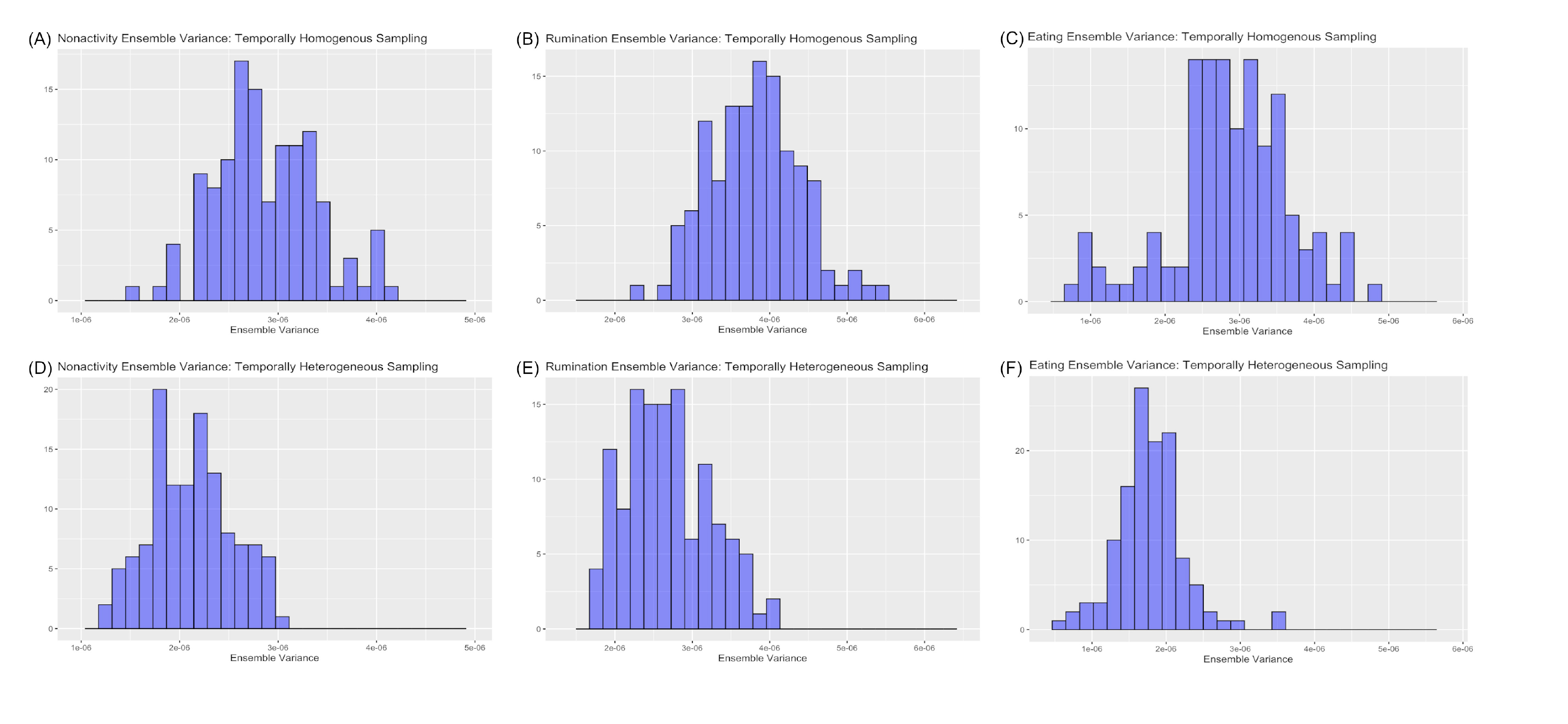}\\
		\caption{Variance comparison on two version mimicked cow-specific ensembles with temporal homogeneity ((A),(B), (C)) and non-homogeneity ((D),(E), (F)) across three behavioral categories: (A, D) non-activity; (B, E) Ruminating; (C,F) eating.}
		\label{fig:cow2}
	\end{figure*}

\begin{figure*}[t]
		\centering
         \includegraphics[width=6in]{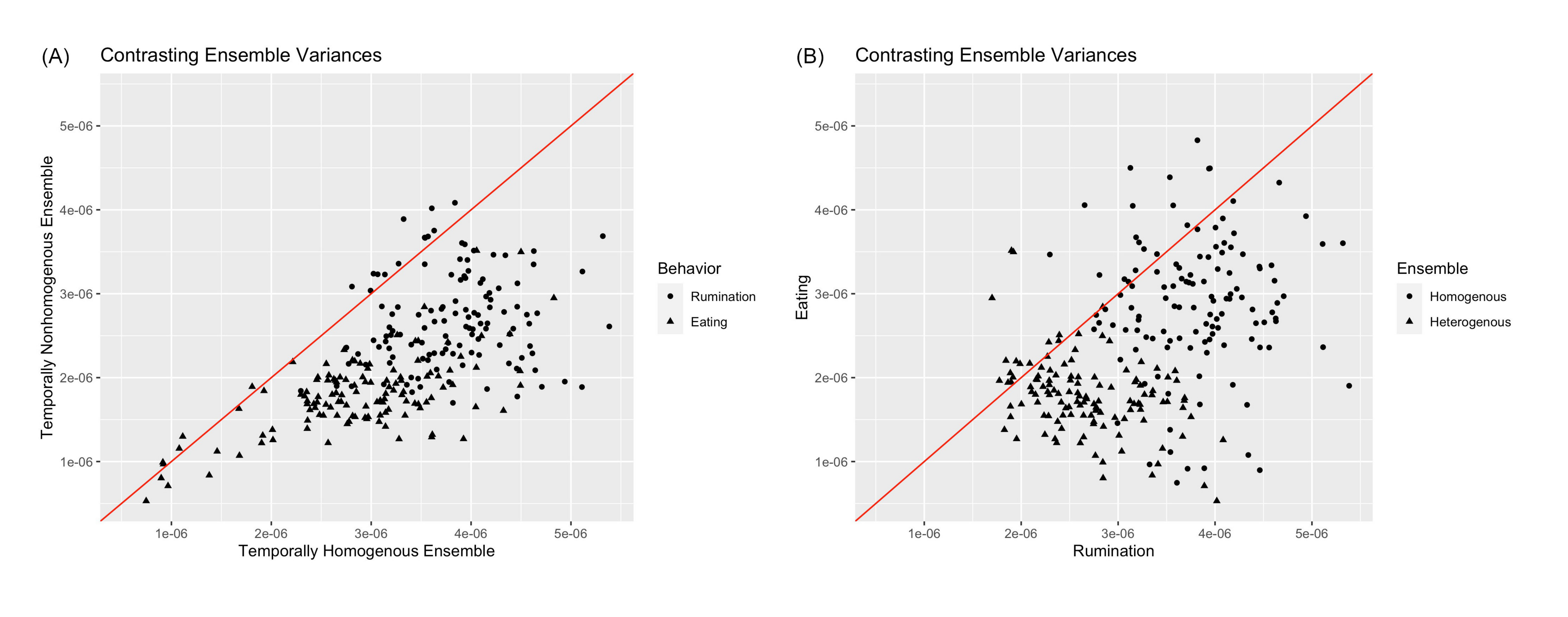}
		\caption{Variance comparison on two version mimicked cow-specific ensembles with temporal homogeneity and two behaviors:(A) homogeneity vs. non-homogeneity w.r.t ruminating and eating, respectively; (B) Ruminating vs. eating w.r.t the two sampling schemes.}
		\label{fig:cow3}
	\end{figure*}

By accommodating the real temporal non-homogeneity, differences on cow-behavior's sample variances are realized. One large scale version of variance comparison is shown in Fig ~\ref{fig:cow2} on three behaviors: nonactive, eating and ruminating. We see that histograms of cow-behavior-specific sample variance across the $B$ matrix mimicries under temporal heterogeneous sampling scheme shift to the left of the histograms under temporal homogeneous sampling scheme. One fine-scale version of variance comparison is given in the panel (A) of Fig. ~\ref{fig:cow3}. We see almost all cows' temporal-heterogeneity-based variances of two behaviors: ruminating and eating, are smaller than temporal-homogeneity-based ones. Further, we also conclude that such a phenomenon of variance reduction indeed occurs simultaneously across all behaviors as illustrated in panel (B) of Fig. ~\ref{fig:cow3}.

\begin{figure*}[t]
		\centering
		\includegraphics[width=3in]{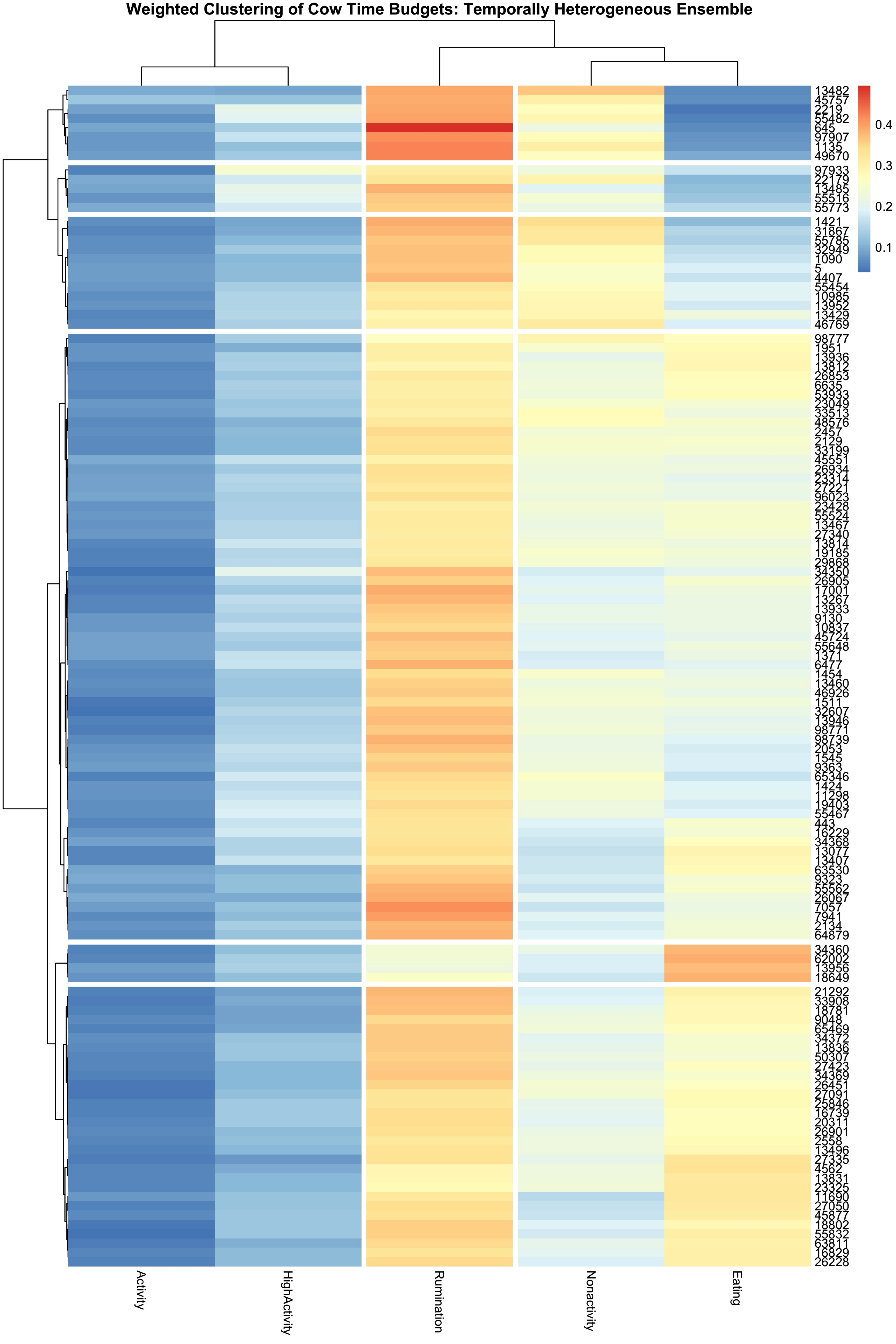}
		\caption{Heatmap with a HC-tree based on a mimicking scheme adopting temporal non-homogeneity and distance measure  $d^*(.,.)$.}
		\label{fig:cow4}
	\end{figure*}

Such variance differences would be factored into the re-scaling via $Var[p^b_{km}]+Var[p^b_{k'm}]$ in defining $d^*(.,.)$ distance measure. The overall effects of being able to truthfully reflect data's authentic stochasticity would be explicitly reflected in the heatmap, as shown in Fig.~\ref{fig:cow4}, which shows more evident uniformness within all column-blocks than the two heatmaps in Fig.~\ref{fig:cow1}. These effects are expected to be translated into the reliability evaluations of all potential observed patterns through the heatmap in Fig.~\ref{fig:cow4}, as would be seen Fig.~\ref{fig:cow5} in the section below . This is the primary merit of adopting truthful stochasticity of data matrix ${\cal P}^o$.

\section{Reliability evaluation}

As each structural pattern is represented by a set of binary code-sequences, evaluating its reliability becomes a relatively simple task. Likewise, we apply the coding scheme on each HC tree $\tilde{T}^b$ computed on mimicry ${\cal P}^b$. Then the most critical parts of information content contained contained in data matrix ${\cal P}^o$ are collected in a two-step fashion: Step-1) Identify a pattern of interest on HC tree $\tilde{T}^o$; Step-2) Evaluate this pattern's reliability based on the clustering-tree ensemble of $\{\tilde{T}^b\}^B_{b=1}$. Such information content, particularly from a group perspective, is invaluable. 

Following these steps, we then want to know what can be attributed to the system's deterministic structures, in contrast with what are largely made out of stochastic structures. Thus, the next key issue arises: how to determine which tree structures are deterministic, which structures are in fact primarily stochastic in nature? That is, we expect heterogeneous reliability or uncertainty attached to any tree's structural component. In this paper, we make use of the ensemble $\{{\cal P}^b\}^B_{b=1}$ to reveal each observed geometric component's reliability. In the next section, we explicitly illustrate what are typical tree structural patterns of interest and how to express them via binary code sequences.

By employing the distance measure $d^*(.,.)$ (modified Euclidean), we obtain a clustering tree ${\tilde T}^o$ on row axis of data matrix ${\cal P}^o$ and $\tilde{T}^b$ on ${\cal P}^b$ with $b=1,.., B$. We seek reliability or uncertainty pertaining to whatever observable structural components of ${\tilde T}^o$. Intuitively, this uncertainty or reliability concept should be based on how often the targeted patterns indeed being conserved within the ensemble $\{\tilde {T}^b\}^B_{b=1}$. We develop our reliability evaluating protocol as follows.

Upon a computed HC tree $\tilde{T}^o$, one binary coding is found by descending from the tree's top internal node to each tree-leaf via a simple binary coding scheme: code-0 for going to left branch and code-1 for the right. Via this coding scheme, memberships of any branch of ${\tilde T}^o$ can be simply identified by a common segment of binary codes shared by all its members' binary codes. Therefore, any observable structural patterns on ${\tilde T}^o$ can be defined via one binary code sequential segment. Two pieces of information are counted from one common code sequential segment: {\bf the length of a binary code segment and the size of tree-leaves sharing the same binary code sequential segment}. These two counts are the two key-ingredients for our reliability evaluation protocol because they exactly determine whether a tree with or without a targeted pattern. We use the protein's Amino Acid content as an example for illustrating purpose.

We select 79 proteins and collect their amino acid content from their sequences. There are 20 amino acids in protein sequences with code-names: \\
$\{R, K, C, W, N, Q, M, F, H, Y, T, P, D, I, E, S, A, G, L, V\}$. A protein sequence, to a great degree, can infer a protein's 3D structure, and consequently imply proteins' spatial or evolutionary closeness or functionally similarity. By completely scrambling a protein sequence, a 20-dim vector of proportions (amino acid contents) seemingly loses all its primary biological information.

Nonetheless, it is still reasonable to ask: whether similarity via amino acid content could offer any biologically relevant information? This question is surely exploratory in nature. That is, from a Data Science perspective and having no a priori knowledge, our data analysis merely aims to provide a small window for biologists to discover what kinds of biologically relevant information could possibly turn out. On the other hand, this categorical data structure with relatively short sequences provides an excellent example of high variability or low reliability.

	\begin{figure*}[t]
		\centering
        \includegraphics[width=4.5in]{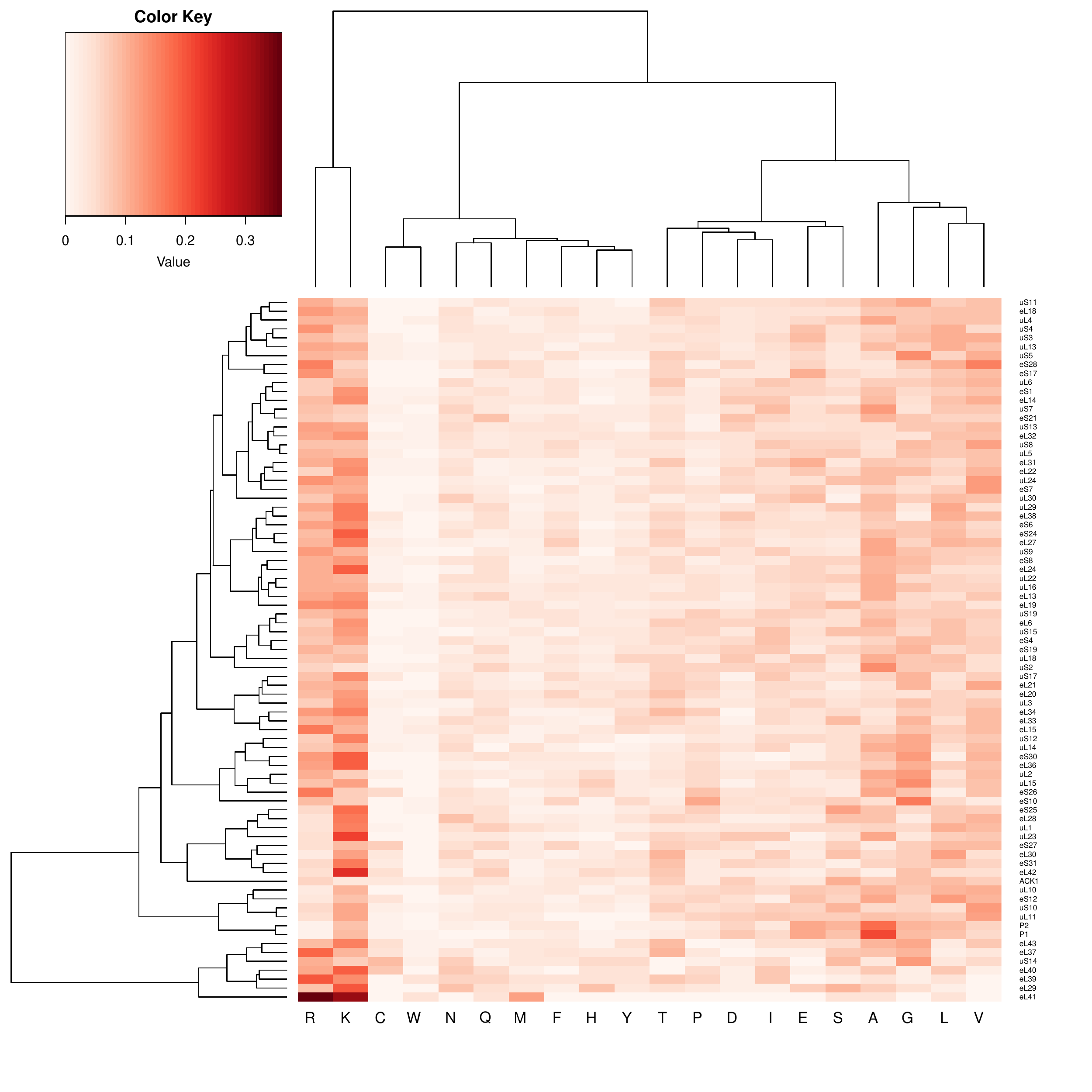}
		\caption{Heatmaps with a HC-tree based on protein data with distance measure $d^*(.,.)$.}
		\label{fig:protein0}
	\end{figure*}

To visualize the data set, the $d^*(.,.)$ distance measure is employed to build a HC-tree and compute a heatmap as shown in Fig. \ref{fig:protein0}. Based on the binary coding scheme: $0$ for Left-branching (lower) and $1$ for right-branching (upper), each protein as a tree-leaf can be located via a binary code sequences within the tree's bifurcating structures. For instance, the bottom tree-leaf is the protein named $[\{eL41\}]^o$. It is attached with a binary code sequence $[00]^o$ with code-length being equal to 2. The 2nd tree-leaf from bottom is the protein named $[\{eL29\}]^o$ with code sequence $[0100]^o$ and code-length 4. The top tree-leaf is the protein named $[\{uS11\}]^o$ with a code sequence $[1111111111111]^o$ of 13 in length.

Further, with a binary code sequence for each tree leaf, an observed branch can be also encoded. For instance, the small branch of 3 proteins $[\{eL29, eL39, eL40\}]^o$ has a binary code sequence $[010]^o$, while a tiny branch of 2 protein $[\{P1, P2\}]^o$ is encoded by $[100]^o$. Very importantly, the idea of common binary code sequence also works for whether the separation of two groups is realistic or not.

Here we illustrate our computational result only on one group of proteins as follows. The four proteins $\{P1, P2, uL10, uL11\}$ are part of a well-known P-stalk region of the ribosome. These four proteins are found sharing a small stand-alone branch (the 2nd from bottom) in the HC-tree derived based on distance measure $d^*(.,.)$, as shown in Fig.~\ref{fig:protein0}. If the a priori knowledge implies these four proteins $\{P1, P2, uL10, uL11\}$ should stay close, then the HC-tree based on distance measure $d^*(.,.)$ seems to rightly capture the pattern.

\begin{figure*}[t]
		\centering
		\includegraphics[width=3.5in]{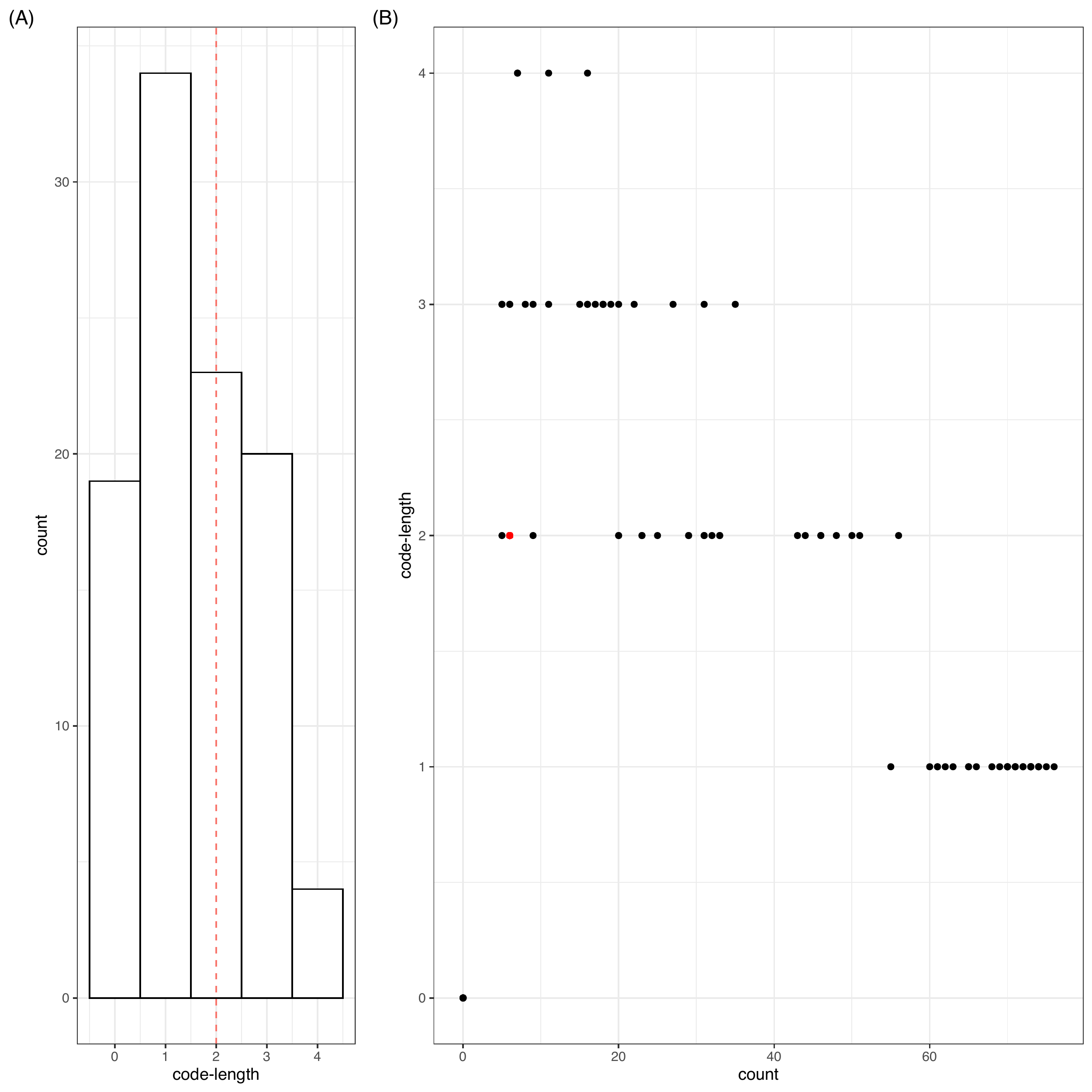}
		\caption{Reliability of 4 proteins$\{P1, P2, uL10, uL11\}$ with $d^*(.,.)$: (A) via coding length;(B) via scatter plots of $100(=B)$ sizes of branches containing these four proteins just before they are separated. The red dot is the observed.}
		\label{fig:protein2}
	\end{figure*}

This pattern of $\{P1, P2, uL10, uL11\}$ on the HC-tree $\tilde{T}^o$ turns out to have relatively small reliability, as shown in Fig.~\ref{fig:protein2}. It is very interesting to note that this seemingly visually evident pattern upon the geometry of $\tilde{T}^o$ is not sustained in the ensemble $\{\tilde{T}^b\}^B_{b=1}$. Only 7 mimicked HC-trees contain these four proteins as in an early bifurcated branch. The protein group is more likely jointed by many other proteins on tree-tops and then gets separated rather early. On the other hand, the group of $\{P1, P2\}$ is a rather robust stand-alone small branch.

Returning to the previous example with behavioral time budgets of dairy cows, we see from the heatmap shown in Fig.~\ref{fig:cow4}, at least two clusters of cows stand out as being visually distinct from the remainder of the herd: the 1st and 5th from top-down direction. The 1st cluster is a group of 8 cows spending a rather large proportion of their time budgets ruminating. In contrast, the 5th cluster is a group of 4 cow spending a rather large proportion of their time budgets eating. That is, cows in these two groups appear to have rather distinct feeding patterns. How reliable are the membership-coupling of these two groups? To answer such a reliability question, we need to represent and define each tree-leaf's coordinate within a tree geometry.

\begin{figure*}[t]
		\centering

		\includegraphics[width=4.5in]{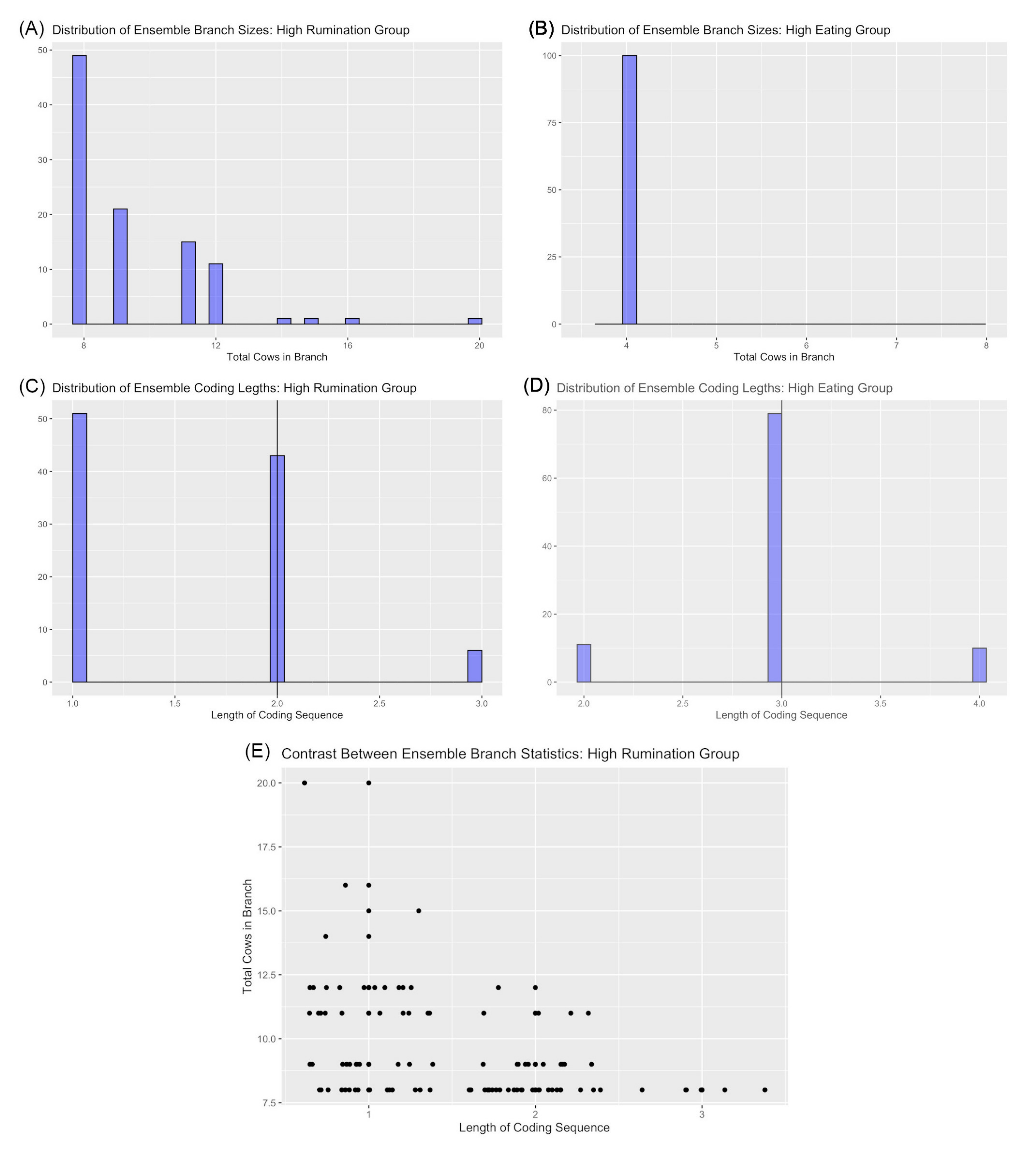}
		\caption{Reliability of 2 groups of cows via code-length and branch-size:(A,C,E) high ruminating; (B,D) high easting, based on a mimicking scheme adopting the temporal non-homogeneity and distance measure  $d^*(.,.)$.}
		\label{fig:cow5}
	\end{figure*}

As seen in panels (B)\&(D) of Fig.~\ref{fig:cow5}, these two groups are rather reliable. The high eating group are found clustered together on a terminal node in all mimicries, diverging from the remainder of the tree within the first 2-4 branches. The high rumination cluster is not as clearly defined, as seen in panels (A)\&(C)\&(E) Fig.~\ref{fig:cow5}. In nearly half the mimicries, they are found together on a terminal node, but in remaining samples their branch contains other animals. In contrasting branch size with code length, we see that in most mimicries these cows are effectively isolated at the second or third branching, but in some samples they are not well distinguished from other animals in the high/low eating distinction made at the first branching. In all mimicries, however, the cows in the high rumination group are separated from the cows in the high eating group at the first branching, suggesting that there exists robust heterogeneity in mastication patterns within this herd that warrants further analysis.

\section{CEDA on other three real examples}
We further analyze a baseball pitchers' pitch-type data set consisting of 747 pitchers, and then one relatively small data set of foreign students' countries of origin in 47 national universities in Taiwan. The contrasting results presented are to be expected. The majority of baseball patterns are relatively reliable, while only a minority of university patterns are.

\subsection{MLB pitchers}
Every pitcher in Major League Baseball (MLB) has a repertoire of pitch-types in order to deal with batters effectively. We collect all pitchers' pitches throughout the entire Year 2017 season. We select those pitchers who pitched more than one thousand pitches. But we exclude 8 pitchers who nearly exclusively pitched ``curveball''. There are 747 pitchers and 15 possible pitch-types. The entire data set is represented in a $747 \times 15$ matrix. Upon this data matrix, we know that a HC-tree based on $d^*(.,.)$ would be more efficient and robust than a tree based on Euclidean distance $d_o(.,.)$. So, we only report the results based on weighted distance $d^*(.,.)$.

	\begin{figure*}[t]
		\centering
		\includegraphics[width=6in]{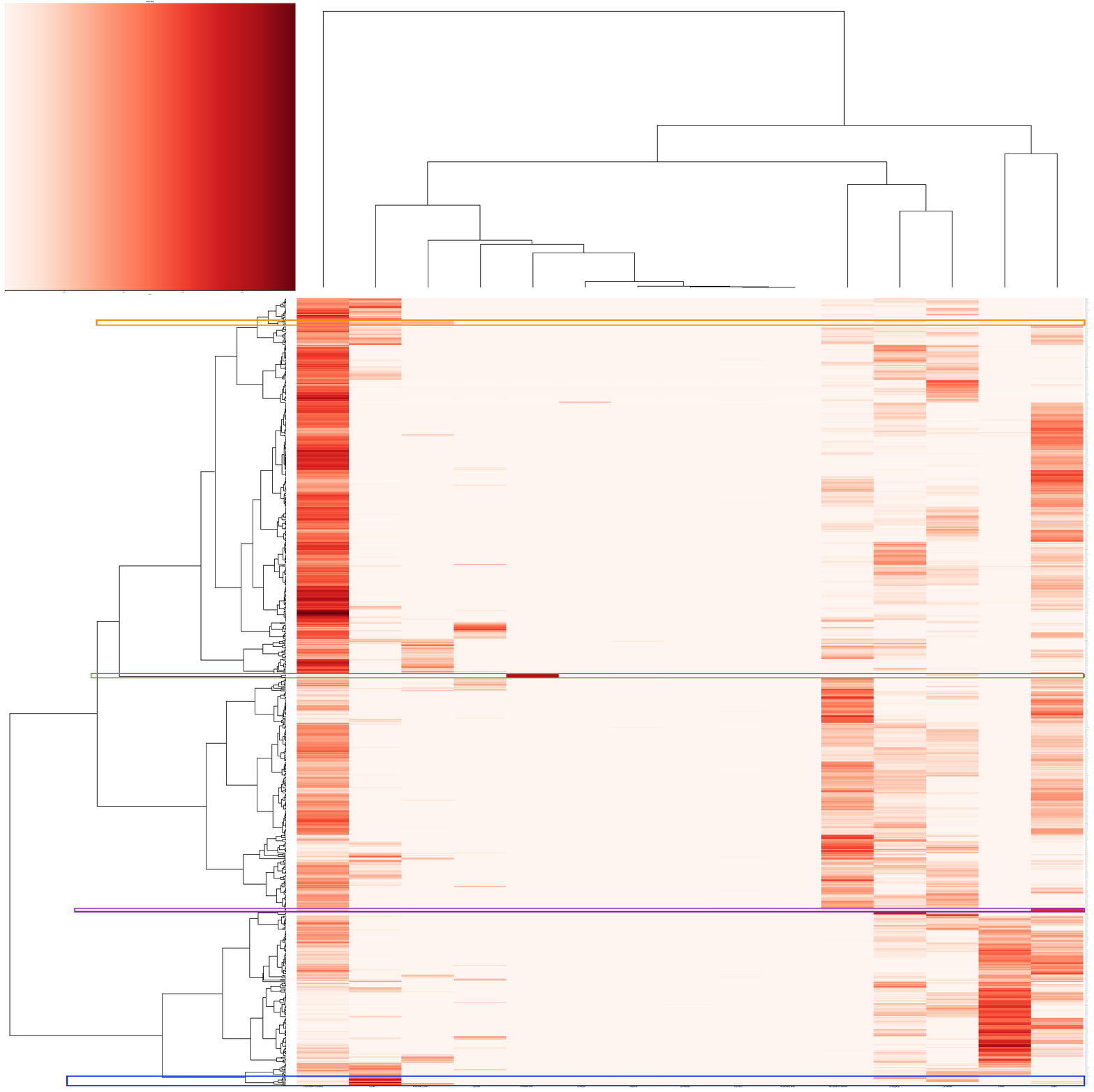}
		\caption{Heatmap of MLB pitchers' pitch-types superimposed with a HC-tree based on distance measure $d^*(.,.)$ among $746(=754-8)$ by excluding 8 pitchers who exclusively pitch curveball.}
		\label{fig:pitcher1}
	\end{figure*}

The heatmap with a superimposed HC-tree on row-axis, as shown in Fig.~\ref{fig:pitcher1}, reveal many characteristically distinct clusters of pitchers. It is evident that a series of patterns can be attached to each pitcher-cluster explain why they are together sharing the same branch, which can be big or small in sizes. We particularly mark two individual pitchers and four small clusters of pitchers, as listed below:
\begin{description}
\item[Individuals]: Clayton Kershaw and Jacob DeGrom
\item[blue box]: Mark Melancon, David Robertson, Kenley Jansen, Mike Bolsinger, Will Harris, Xavier Cedeno, Josh Collmenter, Ryan Merritt;
\item[green box]: R.A. Dickey, Steven Wright, Erick Aybar
\item[orange box]: David Phelps, Wade Davis, Brandon Workman, Brett Cecil, Matt Moore
\item[purple box]:  Jon Jay, Andrew Kittredge, Edward Paredes, Jairo Labourt
\end{description}

\begin{figure*}[t]
		\centering
		\includegraphics[width=5in]{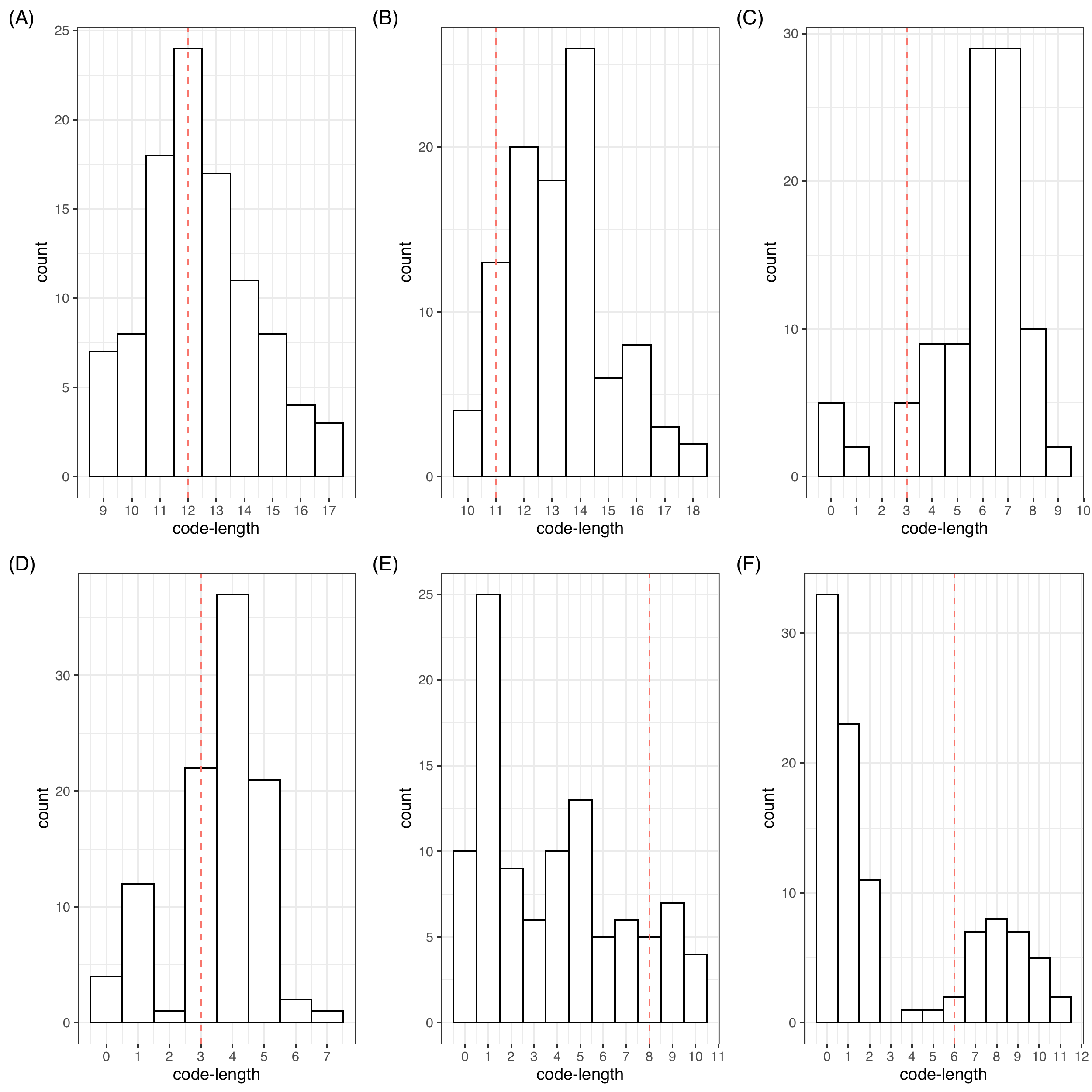}\\
		\caption{Reliability of branches of pitchers.(A)Clayton Kershaw;(B)Jacob DeGrom;(C)blue box; (D)green box; (E)orange box; (F)purple box.}
		\label{fig:pitcher2}
	\end{figure*}

We then evaluate and report these patterns' values of reliability in Fig.~\ref{fig:pitcher2}. From the panels (A) and (B), we can see that these two pitchers are individually similar with many other pitchers in terms of frequencies of pitch-types. So, there are no singular patterns of usages of pitch-types pertaining to two pitchers. Among the four small clusters of pitchers, we found that, except the ``orange'' cluster of pitchers, the rest of three clusters of pitchers are rather robust with high values of reliability.

Certainly, there are many more clusters of pitchers of great interest waiting to be discovered through the heatmap in Fig.~\ref{fig:pitcher1}. In the near future, we plan to set up a website and upload all tables of coding sequences for observed and mimicked HC-trees. We expect that, through interactive Q\&A, readers can go to the web-site to explore all possible patterns and their values of reliability.

\subsection{Home Countries of foreign students at 47 Universities in Taiwan}
We illustrate CEDA through a real data set of country-of-origin of foreign students in national universities of Taiwan. This is one case where row-sums are not large enough.

We collect data of country-of-origin of current foreign students in 47 national universities of Taiwan. These 47 universities are rather heterogeneous in characters. There are only a few complete universities, like NTU, NCKU and NTHU. There are technology oriented ones, like NCTU and NTUST. There are politics and business oriented universities, such as NCCU.  There are university specialized on nursing, like NTIN, on art, like TCPA. We are interested in exploring what kinds interesting associative patterns exist between the heterogeneity in universities and the geographic regions foreign students come from. Through the foreign student body in Taiwan, we have to find relationships between types of universities they attend and the economic interests of their home countries.

\begin{figure*}[t]
		\centering
		\includegraphics[width=3.5in]{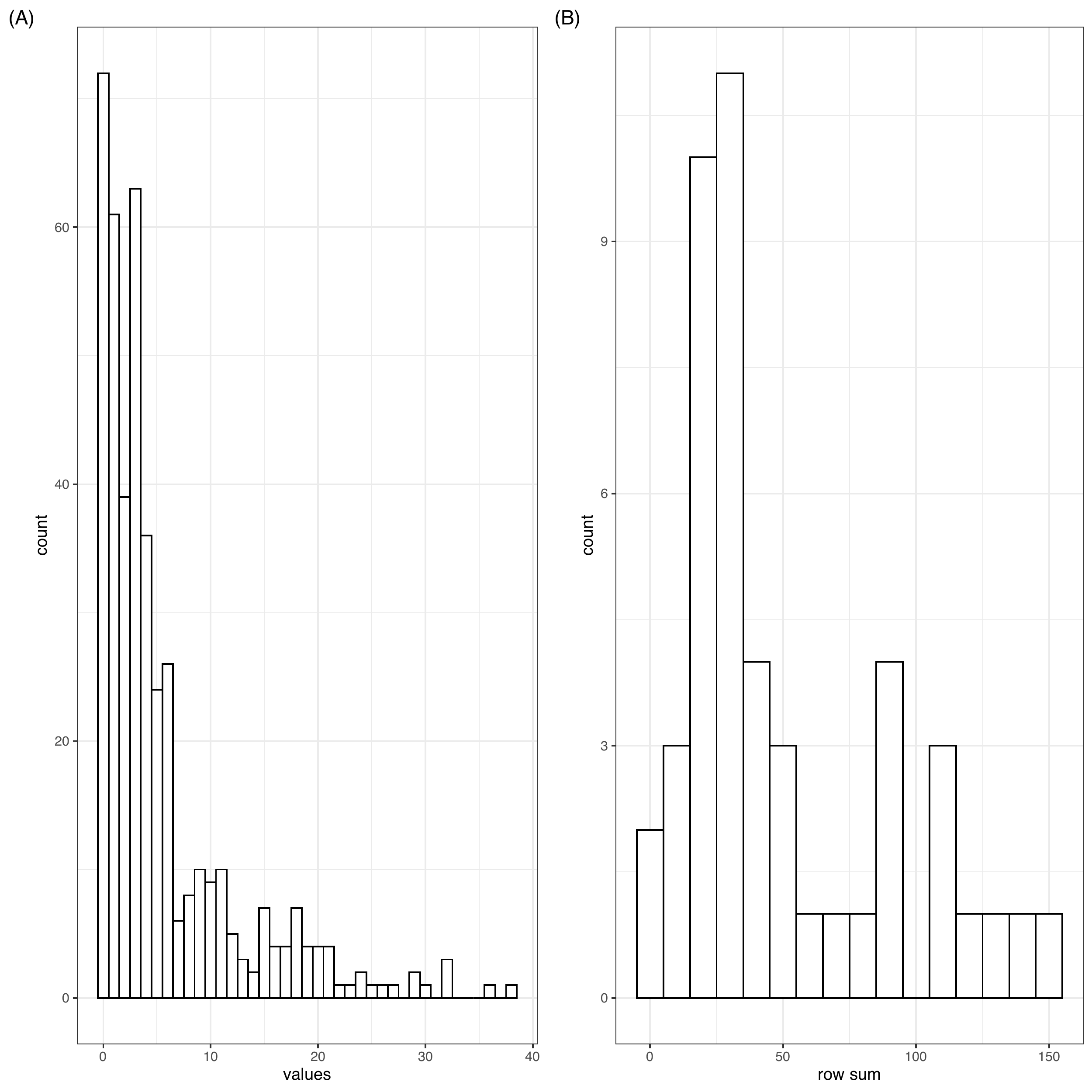}\\
		\caption{Histograms of cell-counts (A) and row sums (B) of $47\times 9$ matrix of foreign students in national universities of Taiwan.}
		\label{fig:school0}
	\end{figure*}

Due to the presence of many small counts across many universities, we group countries into 7 geographic categories: ASEAN, Europe, South Asia, North America, Africa, China, HK/Macau, Other Asia, Central and south America. However, many small counts are still present, as seen in panel (A) of Fig~\ref{fig:school0}. The concern of small counts is essential here because a small count within a university will create a small variance $Var[p^b_{km}]$. When two universities both have small counts within one category they have a high potential of generating large variations in their distance evaluations. That is the case in observed data matrix and even more so in mimicked data matrices. Such a potential is especially high when the row-sums are not large enough, as seen in panel (B) of Fig~\ref{fig:school0}. In fact, this is the case in this data set. It is noteworthy that zero counts in the observed data don't generate or contribute to this issue.

Such large variations will diminish the expected gains by using $d^*(.,.)$ because of instability of distance evaluations. In contrast, the Euclidean distance $d_o(.,.)$ doesn't suffer from such an issue created by small counts. Therefore, the differences between these two distance measures used in constructing HC-trees are evident through various patterns' reliability, as would be seen below.

\begin{figure*}[t]
		\centering
		\includegraphics[width=0.45\textwidth]{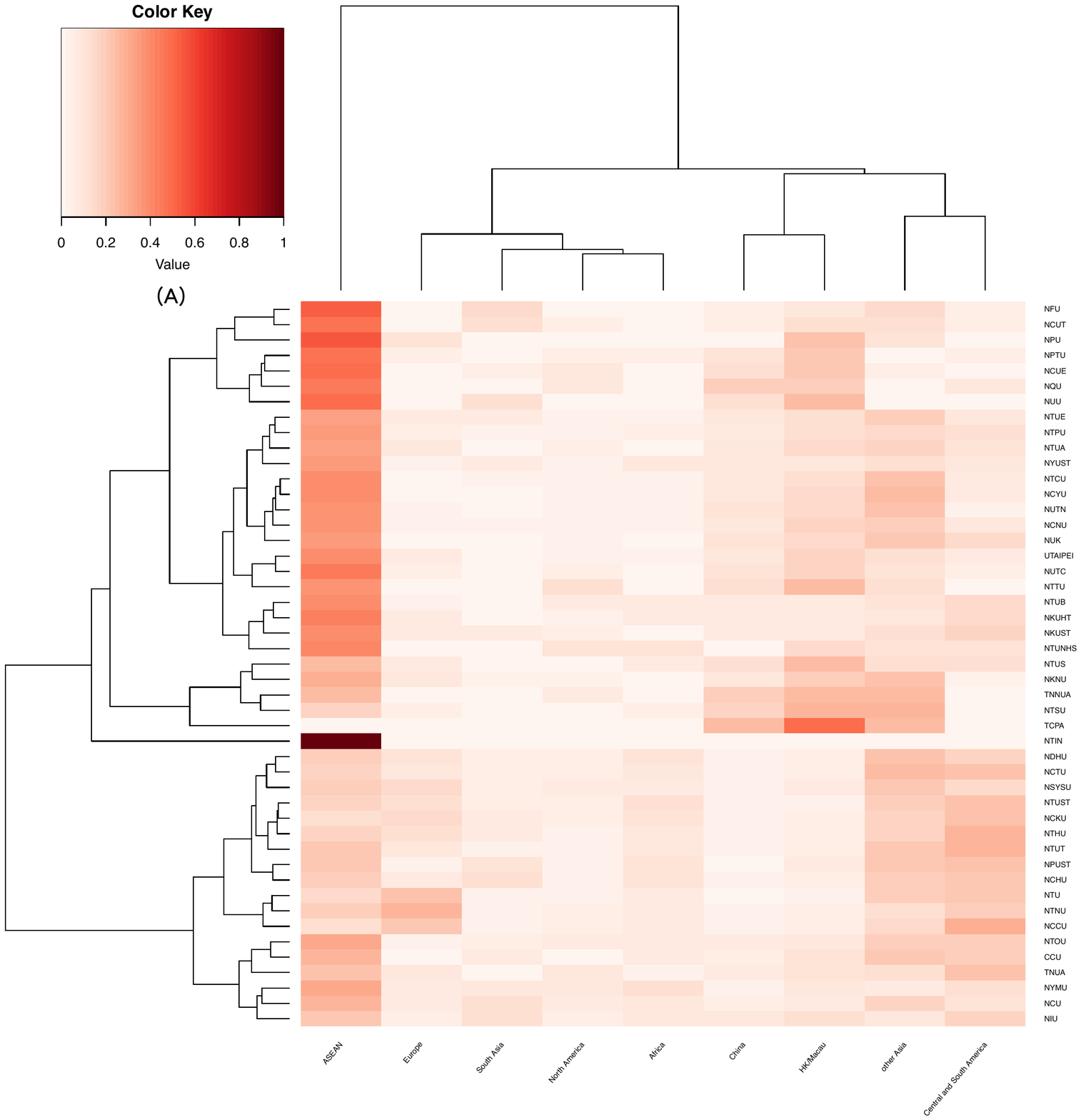}
		\includegraphics[width=0.45\textwidth]{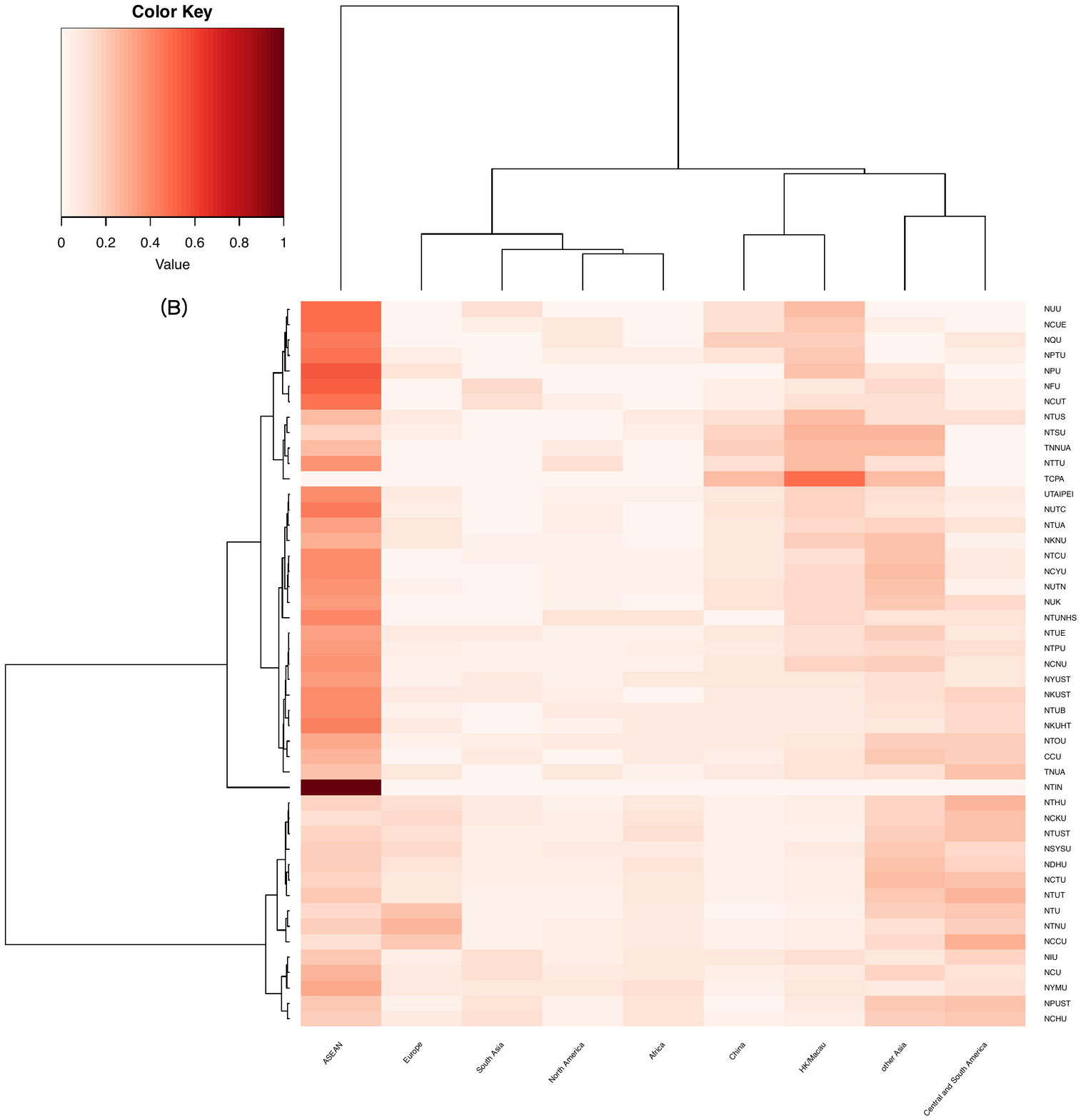}
		\caption{Heatmaps of foreign students in Taiwanese universities superimposed with a HC-tree based on distance measure:(A)  $d_o(.,.)$; (B) $d^*(.,.)$.}
		\label{fig:school1}
	\end{figure*}

The association between geographic region and university character can more or less be seen through the heatmaps in Fig.~\ref{fig:school1}. We see European students are in three major universities, while students from Chain and HK/Macau are attending relatively smaller universities located on the upper branch. Despite the difference in distance range, it is evident that the two HC-trees are rather similar in branching on large as well as on local scales. For instance, the singleton branch $\{NTIN\}$ and two small branches: $\{NTU, NTNU, NCCU\}$ and $\{NTHU, NCKU\}$, are found in relative similar locations of the two trees.

However, the reliability of the two small branches: $\{NTU, NTNU, NCCU\}$ and $\{NTHU, NCKU\}$ are different. The Euclidean distance $d_o(.,.)$ offers a slightly larger reliability value for the tiny branch $\{NTHU, NCKU\}$ then $d^*(.,.)$ does, while offers significantly larger reliability value for $\{NTU, NTNU, NCCU\}$ than $d^*(.,.)$ does. It is a bit surprising that reliability evaluations based on both distance measures strongly indicate that the observed singleton branch $\{NTIN\}$ is not that reliable.

	\begin{figure*}[t]
		\centering
		\includegraphics[width=4.5in]{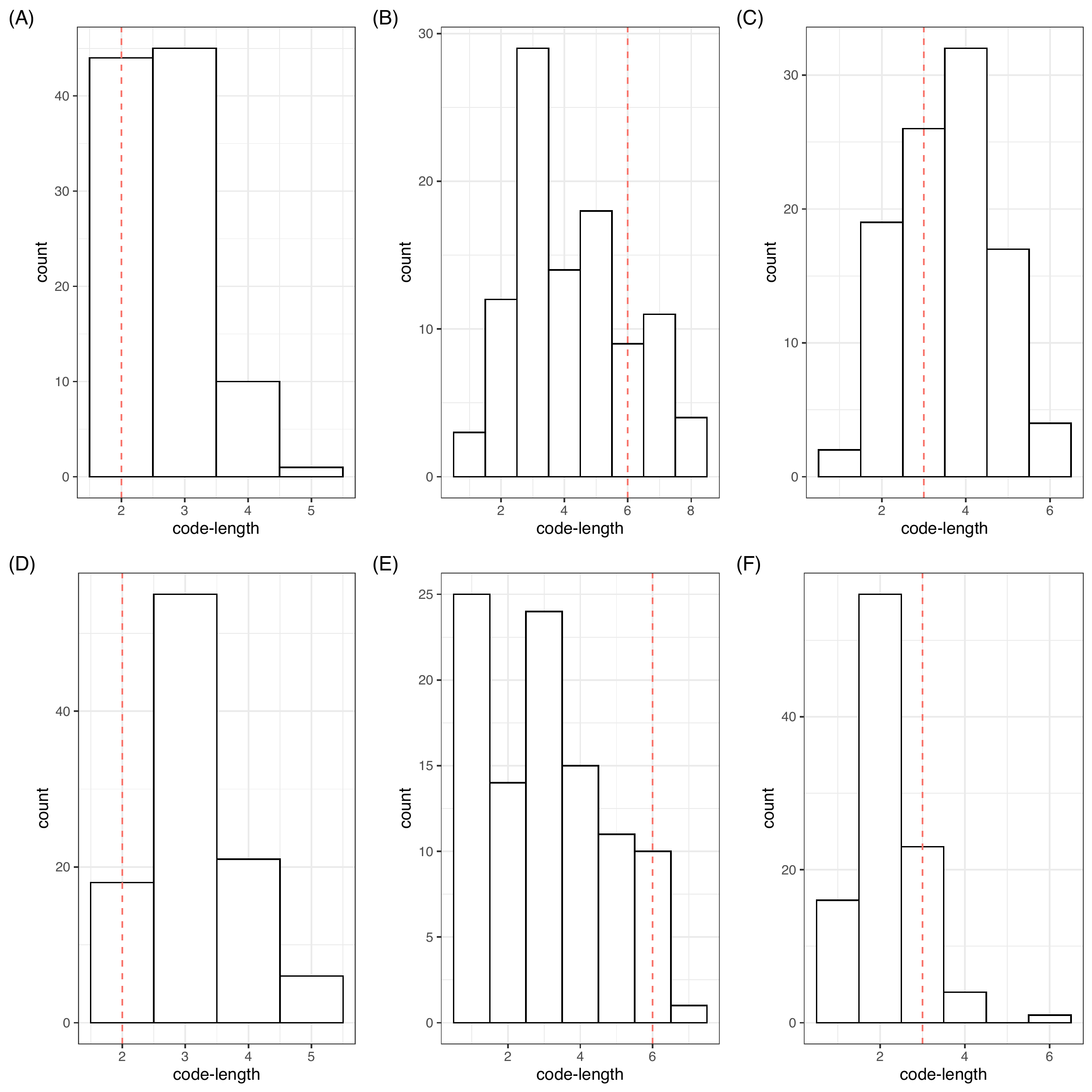}
		\caption{Reliability of 3 patterns on two HC-tree based on distance measure based on $d_o(.,.)$ ((A), (B), (C) and $d^*(.,.)$ ((D), (E), (F)): ((A), (D)} $\{NTIN\}$; ((B),(E)) $\{NTHU, NCKU\}$; ((C),(F))) $\{NTU, NCCU, NTNU\}$
		\label{fig:school2}
	\end{figure*}

\section{Conclusions.}
For comparing a large number of populations via a categorical variable, as termed Extreme-$K$ categorical-sample problem here, is rather a common problem we encounter in real world. But it is strange we seldom research about it. Beyond those issues we discussed here, there many issues remain waiting to be solved. Such as, how to accommodate randomness from bins' boundaries of a histogram under the continuous measurement setting. However we refrain from going to that in this conclusion section. Here we briefly reiterate key concepts discussed in the previous sections under Extreme-$K$ categorical-sample problem.

Our focus are placed on patterns of histograms' global shapes. With a large $K$, the number of all possible population-pairs is $\frac{K(K-1)}{2}$ already very big. If we go beyond the pairwise comparisons, then the number of potential comparisons would be Exponentially big. That is to say that a priori determined comparisons are out of practical considerations. On the other hand, we need to present computable data's information content in a fashion that facilitates all potentially viable comparisons and their reliability-evaluations.

Our global scale comparisons are practically phrased as issues of ``which populations' histogram-shapes are close to which populations', but far away from which populations' '', then resolutions of such global issues must be resolved by constructing a geometry upon the observed data matrix ${\cal P}^o$.
Specifically, we propose a proper distance measure and build a tree geometry on the population space. So that whoever is interested on such an extreme-$K$ categorical samples can discover patterns of interest and their reliability.  This is the chief objective of our CEDA developments.

Our CEDA enables us to achieve the following goals with computational effectiveness: 1) handle a framework that Analysis of Variance (ANOVA) in Statistics cannot handle; 2) build a tree geometry as a discovery platform for a wide spectrum of objectives embedded within $K$ populations; 3) evaluate uncertainty or reliability of all objective via mimicking. The huge number of group-based comparisons as our collective of objectives inspires a natural link to artificial intelligence (A.I.). We argue that, with a properly chosen or constructed distance measure, a clustering based tree geometry can be built as a population embedding tree, which serves as a natural platform for discovering viable and pertinent objectives hidden and embraced by such a data framework. Thus, such a tree geometry can effectively exhibit complexity of data's information content. All global and local geometric structural patterns (in various forms of group-comparisons) pertaining to such a tree are mathematically represented via binary coding sequences of all tree-leaves (population-IDs). Based on the ensemble of matrix mimicries, we are able to evaluate each geometric structural pattern's computable reliability or uncertainty. We conclude that the data's information content is explicitly visible and readable via a population embedding tree geometry and its heatmap coupled with uncertainty heterogeneity.

From a machine learning (ML) perspective, though this Extreme-K categorical-sample problem is a special Multiclass Classification setting by having one single feature, all machine learning (ML) techniques will not be efficient. Furthermore ML methodologies all aim at predictive capability, but there is not much room for prediction here. These two facts might explain to a great extent why this setting is not popularly researched. We emphasize again that extracting data's full information content as the objective of data analysis is far beyond the inferential purpose.


\bibliographystyle{unsrt}

\end{document}